\documentclass[12pt,fleqn,notitlepage]{article}
\usepackage[authoryear,sort&compress]{natbib}
\usepackage{graphicx}
\usepackage{latexsym}
\usepackage{amsfonts}
\usepackage{amsthm}
\usepackage{epsfig}
\usepackage{mathrsfs}
\usepackage{epic}
\usepackage[leqno]{amsmath}
\usepackage{amssymb}

\usepackage{setspace}
\newtheorem{lemma}{Lemma}[section]
\newtheorem{theorem}{Theorem}

\newcommand{\Hh}{\mathcal H}

\newcommand{\Pc}{\mathcal P}

\newcommand{\Rb}{\mathbb R}

\begin{document}
\raggedright

\begin{center}
\noindent \textbf{Species, Clusters and the `Tree of Life':\\
\noindent A graph-theoretic perspective}\\

\bigskip

\bigskip
\noindent ANDREAS DRESS\\
\noindent CAS-MPG Partner Institute for Computational Biology, 320 Yue Yang Road, 200031 Shanghai, China; \\E-mail: andreas@picb.ac.cn\\

\bigskip
\noindent VINCENT MOULTON\\
\noindent School of Computing Sciences, University of East Anglia, Norwich, NR4 7TJ, UK; E-mail: v.moulton@uea.ac.uk\\
\bigskip
\noindent MIKE STEEL\\
\noindent Allan Wilson Centre for Molecular Ecology and Evolution, Biomathematics Research Centre, University of Canterbury, Christchurch, New Zealand;\\ E-mail: m.steel@math.canterbury.ac.nz\\

\bigskip
\noindent TAOYANG WU\\
\noindent CAS-MPG Partner Institute for Computational Biology, 320 Yue Yang Road, 200031 Shanghai, China;\\
\noindent School of Computing Sciences, University of East Anglia, Norwich, NR4 7TJ, UK; \\E-mail: taoyang.wu@uea.ac.uk\\

\bigskip

\end{center}

\begin{abstract}
\noindent A hierarchical structure describing the inter-relationships of species has long been a fundamental concept in systematic biology, from Linnean classification through to the more recent quest for a `Tree of Life.'   In this paper we use an approach based on discrete mathematics to address a basic question:  Could one delineate this hierarchical structure in nature purely by reference to the `genealogy' of present-day individuals, which describes how they are related with one another by ancestry through a continuous line of descent? We describe several mathematically precise ways by which one can naturally define collections of subsets of present day individuals so that these subsets are nested (and so form a tree) based purely on the directed graph that describes the ancestry of these individuals.  We also explore the relationship between these and related clustering constructions.

\end{abstract}

Keywords: species, ancestry, hierarchy, cluster, digraph
\newpage

\section{Introduction}

In this paper, we apply discrete mathematical arguments to study how hierarchical structures arise naturally from a very basic graph in systematic biology.

Consider the collection of all organisms that ever lived on earth -- this includes not just the set $X$ of organism
alive at present, and other organisms we can directly observe (e.g. fossil specimens), but a much larger set $V$ consisting of all organisms (or vertebrates or dicots or ...) that ever lived on this planet.
There is a very natural directed graph structure on $V$: place a directed arc from $u \in V$ to
$v \in V$ if $u$ was a `parent' of $v$.  Here, the word `parent' means that $u$ contributed directly to the genetic make-up of $v$;  in a sexually-reproducing population, this is the usual meaning of the word
(the two parents of $v$ are the contributors of the sperm and egg), while in an asexually reproducing (haploid) population, each individual typically has one parent (e.g. the prokaryote cell whose division led to the new cell) though,  occasionally, $v$ may be regarded as having additional `parents' beyond those described, as a result of processes such as lateral gene transfer  (LGT) or  other forms of reticulate evolution (e.g. a hybrid taxa).

This graph -- let us call it $G$ -- can thus be regarded as a `history  of life' network, that describes how different past and present individual organisms are related to one another by ancestry
\citep{steel}.  The graph $G$ cannot be directly observed --
we have access only to a subset $X$ of $V$ of `observable' individuals
along with some clues as to the gross structure of the rest of the graph gleaned from the genomic data of individuals in $X$,
and other observable information (morphology, biochemistry, behavior, fossils etc).  Nevertheless, the graph $G$ is a well-defined entity,
based on the premise that each organism has at least one parent, back to the earliest forms of life that existed on earth.

Such a huge graph would not be of much interest were it not for Darwinian evolution. The idea that all life traces back to one common ancestor suggests that $G$ is a connected graph, with the lines of descent of populations that we call `species' merging (coalescing) as we trace their ancestry, from child to parent, backward in time.  Thus, rather than being an isolated set of component graphs -- one for each `species'  -- the graph $G$ is more like a very large, diffuse `tree of populations' (see Fig.\,1), where the populations occasionally split when a `speciation event' occurs, for example when a population becomes separated into two reproductively isolated groups (a process referred to as allopatric speciation), though occasionally these lineages may later intersect, for example if hybrid species arise from two lineages. At the microbial level, with extensive LGT, and occasional endosymbiotic events, this picture may appear more like a `net of life' \citep{koo}.

\begin{figure}[ht]
\label{figure0}\begin{center}
\resizebox{8cm}{!}{
\includegraphics{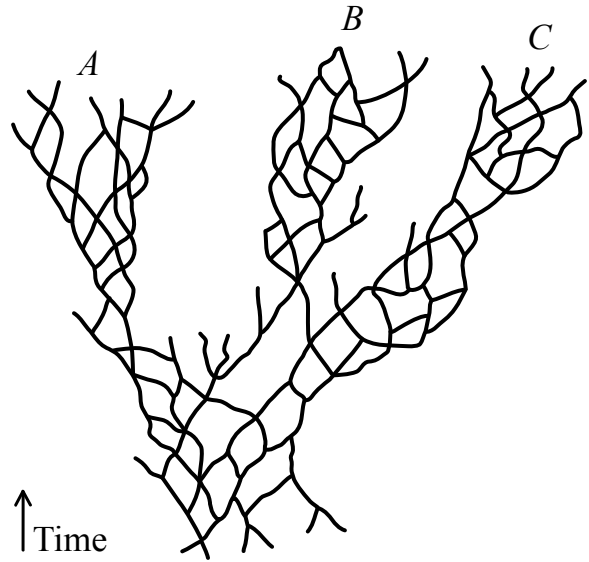}
}
\end{center}
\caption{A simplified picture of a history of populations.  In this example $A,B$ and $C$ form tight clusters.}
\end{figure}

The history of populations is usually represented in systematic biology as  a rooted {\em phylogenetic tree} -- that is a rooted  tree where the leaves are labeled by the extant `species', and which has edges and interior vertices that correspond to ancestral `species' and `speciation events', respectively  \citep{sem, fel}.  In this representation, the fine detail of the descent of a population through time is lost, creating an unfortunate separation between phylogenetics and population genetics.

This high level picture of evolution via phylogenetic trees is problematic  for two further reasons. Firstly, it requires one to address the much-debated notion of the nature and definition of `species', a concept that is particularly ambiguous at the microorganism level \citep{doo, whe}.  Secondly, it is increasingly being argued that processes of reticulate evolution such as LGT require that the evolution of `species' should really be described by a network rather than a tree \citep{doo, koo, dag2, law}.

 In this paper, we take a simple if somewhat novel approach to this issue by asking whether we can simply use $G$ directly to define a tree (or tree-like structure) that reflects
 the bifurcating history of life studied in evolutionary theory, and which (i) does not require the prior identification or definition of `species' and (ii) is robust to the many processes that can complicate a tree-like history, such as LGT.  Viewing an evolutionary tree in this direct way is perhaps in the spirit of Darwin's suggestion to ``discover and trace the many diverging lines of descent in our natural genealogies'' \citep{dar}. Of course, the notion that there is a hierarchical structure to the life we see today is a concept that came well before Darwinian evolution; for example, Linnean classication \citep{lin}  dates back more than 100 years before Charles Darwin's {\em On the Origin of Species by Means of Natural Selection, or the Preservation of Favoured Races in the Struggle for Life} appeared.  Moreover, the nature of `species'  has been discussed much earlier -- from Plato through to the 17th Century English naturalist John Ray.

 In this paper, we do not  provide any general procedure for constructing hierarchies from genomic data; our interest here is purely in addressing the more fundamental questions:
 \begin{itemize}
 \item
Can we construct from $G$ systems of clusters (subsets of $X$) that reflect complex ancestral relationships and yet behave in a nested (tree-like) fashion?
 \item
What are the properties of, and relationships between, different possible constructions?
 \item
 What assumptions, if any, concerning evolution are required so that the clusters derived from $G$ are guaranteed to form a tree?
 \end{itemize}

Fortunately, for this last question, we can be confident about one very helpful property: $G$ has no directed cycles, simply because a `parent'
is always born before its child. We ask then whether any acyclic digraph $G$ with a distinguished subset $X$ of its vertex set
induces a natural rooted tree structure on $X$ (described in terms of a hierarchy, i.e. a system of nested subsets of $X$) that
reflects the process of populations splitting and separating through time. We describe several ways to define such hierarchies,
and we explore their properties and the connections between them.

The use of discrete mathematics to investigate possible tree-like systems of classification arising
in evolutionary biology  more systematically has been explored by a number of authors from different perspectives. For example, \cite{ald} recently considered three  formal ways whereby genera could be defined in terms of species, based on a phylogenetic tree,
 obtaining an elegant characterization of these three classifications (Theorem 1 of \cite{ald}).  A number of authors in the edited volume \citep{mir} deal
 with the  mathematical aspects of defining hierarchies and related structures in biology.  However, all these approaches to date have worked at
 a level that is `higher' than $G$.

Our approach combines two themes developed in our earlier (independent) investigations into processes whereby trees arise by general connectivity considerations in two situations: (i) a general setting of locally connected topological spaces  \citep{dre}, and (ii) a particular metric space associated with  ancestry within populations \citep{steel}.

The structure of the paper is as follows.  We begin by introducing some further definitions, followed by some comments concerning a purely `genetic' variant of the graph $G$. We will define five general ways of obtaining a collection of subsets of $X$ from $G$ based on notions of ancestry.
Our main result (Theorem~\ref{mainthm}) asserts that these all lead to hierarchies (or a related structure, a weak hierarchy), and describes some connections between them. In the final section, we explore some properties of these constructions further.

\section{{\sc Notation}}
Consider a  finite, directed, and cycle-free graph (i.e. an acyclic digraph)
$G=(V,E\subseteq V\times V)$ and the associated partial order ``$\preceq$'' = ``$\preceq_G$'' of $V$ defined, for all $u,v\in V$, by $u \preceq v$ if and only if there exists a (directed) path from $u$ to $v$ in $G$, i.e., a sequence $u_0:=u,u_1,\cdots,u_k:=v$ of some length $k\ge 0$ of elements in $V$ with $(u_{i-1},u_i)\in E$ for all $i=1,\hdots,k$ in which case $u$ will also be called an {\em ancestor} of $v$, and $v$ a {\em descendant} of $u$. Note that we also write $u\prec v$ in the case where
$u\preceq v$ and $u\not =v$ holds.

We will sometimes
refer to the elements of $V$ as
 {\em individuals} and, given any arrow $(u,v)$ in $E$,
the individual $u$ will be called a {\em parent} of $v$ and $v$ will be  a {\em child} of $u$.
Clearly, given any two elements $u,v$ in $V$, we have $(u,v) \in E$ if and only if $\#\{w\in V: u \preceq w \preceq v\}=2$ holds.

Let $X$ denote a distinguished subset of $V$, which we will regard as a set of `observable individuals' in $G$ (e.g. present-day individuals, and perhaps some fossil specimens).
While  no specific conditions need to be placed on $X$ in what follows, it may be natural to assume that every $v$ in $V-X$ has a descendant in $X$ (implying  in particular  that $X$ contains all elements $v\in V$ that do not (yet) have
any children), as eliminating all elements from $V-X$ that do not have a descendant in $X$ will not change the clusters in $X$ we are going to consider below.

For any $v \in V$, let $\overrightarrow{v}=\overrightarrow{v_{_X}}$ denote the set of individuals in $X$ that are descendants of $v$, and for any subset  $U$ of  $V$, put $\overrightarrow{U}:=\bigcup_{v\in U}\
\overrightarrow{v}$.

\begin{figure}[ht] \begin{center}
\resizebox{8cm}{!}{
\includegraphics{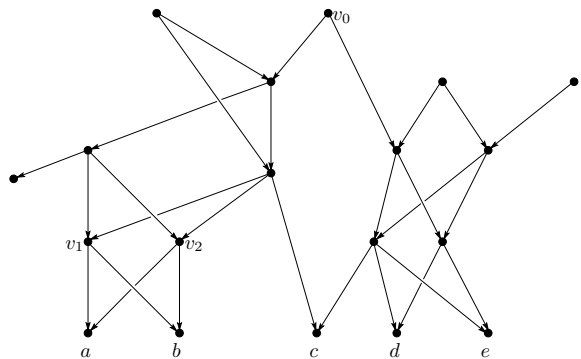}
} \caption{An illustrative  example of an acyclic digraph $G$, with vertex set $V$ and $X=\{a,b,c,d,e\} \subset V.$}
\end{center}
\label{motiv}
\end{figure}

\subsection{Organismal history versus genetic history}

The graph $G$ we have defined in the introduction describes the detailed genealogical history of individual organisms.  However it may also be of interest to consider a subgraph of this graph that reflects just those lines of descent that carry genetic material that survives in at least one of the organisms in our observed set $X$.  Clearly it is possible for an individual organism that lived long ago in a diploid population to have many descendants today, and yet have no surviving genetic material (gene, homologous nucleotide, etc) today due to the processes of population genetics (a gene is inherited from one parent, not both).  This distinction between genetic ancestry and organismic ancestry has been noted by many authors over the years, and has been discussed recently by  \cite{bau}, and, more theoretically, by \cite{mat}.

We can formalize this distinction as follows: let us say that an arc $(u,v)$ of $G$ is (genetically) {\em trivial} if none of the genome of $v$ that is inherited from $u$ is present in any of the descendants of
$v$ in $X$.    Let $G_g$ be the graph obtained from $G$ by deleting all the genetically trivial arcs.  Thus, in $G_g$ we only retain those parent-child arcs for which the child inherits from that parent genetic material that survives in at least one of the observed individuals.

Many of our results (including our main result, Theorem~\ref{mainthm}) remain true for both types of graphs, since they are stated in the generality of a finite, directed, cycle-free graph that contains
$X$ within its vertex set, and $G_g$ clearly inherits these properties from $G$. However, some examples (eg. the example of a tight cluster involving humans), and some discussion depends more crucially on
which type of graph we are considering, and so, for the sake of simplicity, we will regard $G$ as the genealogical rather than ancestral genetic graph from now on.

\subsection{Hierarchies and weak hierarchies}
We say that a collection $\Hh$ of subsets of $X$ forms a (generalized) {\em hierarchy} on $X$ if $\Hh$ satisfies the nesting property:
$$A,B \in \Hh \Rightarrow A \cap B \in \{\emptyset, A, B\}.$$
Note that this condition is also referred to in the hypergraph literature as a {\em laminar family}, and the word `hierarchy' often also requires
further
conditions such as $X \in \Hh$, $\emptyset \not\in \Hh$,  or $\{x\} \in X$ for all $x \in X$.  Here, however, we will insist on the nesting property, only.

A natural bijection exists between (isomorphism classes of) rooted $X$--forests and hierarchies on $X$ that do not contain the empty set (see, for example, \cite{ed}, Section 8) which restricts to a bijection between the set of (isomorphism classes of) rooted $X$--trees and the set of hierarchies on $X$ that contain $X$ but not $\emptyset$. In particular, $|\Hh| \leq 2|X|$ holds for every hierarchy $\Hh$ (maximal hierarchies are considered further by \cite{boc2}).
Note also that if $\Hh$ is a hierarchy on $X$, then so is any subset of $\Hh$, and also that, for any set $Y \subset X$, the collection
$\{A \cap Y: A \in \Hh\}$ is a hierarchy on $Y$.
Given any collection $\Pc$ of subsets of $X$, there is a simple way to define an associated hierarchy $ \Hh_\Pc$ by setting:
\begin{equation}
\label{forfree}
\Hh_\Pc:=\big\{C\in \Pc: \forall  C' \in \Pc, C \cap C' \in \{C,C',\emptyset\}\big\}.
\end{equation}

A weaker condition than that satisfied by a hierarchy is the condition:
$$A,B, C  \in \Hh \Rightarrow A \cap B \cap C  \in \{A \cap B, B \cap C, A \cap C\}.$$
If $\Hh$ satisfies this condition, it is said to form a {\em weak hierarchy}. Weak hierarchies share some properties with `proper' hierarchies (for example, clusters can be identified using at most two elements from $X$), and these are explored further by \cite{bandelt}; moreover, as with a hierarchy, there is a polynomial bound on the size of a weak hierarchy in terms of $|X|$: We have
$|\Hh| \leq \binom{|X|+1}{2}$ for any weak hierarchy that does not contain the empty set.

\subsection{Connectivity through evolution}

Evolution suggests that all organisms we can observe today descended from a small group of common ancestors and this suggests that the graph $G$ is connected in various possible ways. These are summarized by the following, increasingly liberal connectivity requirements:

\begin{itemize}
\item[(C1)] $G$ contains a vertex $v$ with $\overrightarrow{v} = X$.
\item[(C2)] For all $x,y \in X$, there exists $v \in V$ with $v \preceq x,y$.
\item[(C3)] The graph $\Gamma(X): = (X,  \big\{\{x,y\}\in \binom{X}{2}: \exists v\in V: x,y\in \overrightarrow v\big\})$ is connected.
\end{itemize}

In the biological context, Condition (C1)  is merely the statement that all living organisms today have (at least) one common ancestor some time in the past.
Condition (C2) says that every pair of individuals in $X$ has a common ancestor, while Condition (C3) says any two individuals in $X$
are related through a chain of relatives in $X$.
Mathematically, (C2) implies that $\Gamma(X)$ is a complete graph; moreover, we have (C1) $\Rightarrow$ (C2) $\Rightarrow$ (C3).
Although (C1) is usually held to be biologically reasonable \citep{cri, fut, woe, sob}, we do not necessarily assume this condition here; the choice of any particular condition (C1)--(C3) is relevant only for two reasons: (i) It can determine whether or not $X$ is an element of
some of the hierarchies we construct and (ii) Condition (C2) can be helpful to ensure the existence of clusters defined by pairwise ancestral relationships.

\section{{\sc $X$--Clusters from $G$}}

We now describe a variety of ways whereby an acyclic digraph $G$ with $X \subseteq V$ can naturally give rise to specific collections of subsets of  $X$ based on concepts of ancestry. In Section 4, we will show how these constructions lead to (weak) hierarchies.

\subsection{Tight clusters}

We begin with an intuitively simple way to generate clusters on $X$ from any acyclic digraph $G=(V,E)$ with $X \subseteq V$. Although the conditions a cluster must satisfy in this first definition are more severe than 
those we consider later, we will describe in the 
remark below how results in population genetics provide some justification for the existence of such tightly-constrained clusters.

For a non-empty subset $C$ of $X$, let $D(\!\supseteq\! C)$ denote
the set of all individuals $v\in V$
whose descendants contains every individual in $C$,
let
$D(\!\subseteq\!C)$ denote the set of individuals in $V$
all of whose descendants in $X$ are contained in $C$,
and let $D(\!=\!C):=D(\!\supseteq\!C)\cap D(\!\subseteq\!C)$  denote the set of all individuals in $V$ whose
descendants in $X$ coincides exactly with $C$. That is, we put:
$$
D(\!\supseteq\!C)
:=
\{v \in V: \overrightarrow{v} \supseteq C\},\quad
D(\!\subseteq\!C):=
\{v \in V: \overrightarrow{v} \subseteq C\},
$$
and put:
$$
D(\!=\!C):=
\{v \in V: \overrightarrow{v} = C\},
$$
So, $D(\!=\!C)$ consists of all individuals in $V$
that are ancestral exactly to {\em every} element in $C$, but no other elements in $X$.

We define a subset $C$ of $X$ to be a
{\em tight} cluster (in $X$ relative to $G$) if and only if it is non-empty and $D(\!=\!C)$ separates $C$ from $X-C$, that is, every (undirected) path from an element in $C$ to an element in $X-C$ contains some element from
$D(\!=\!C)$.

Note that for any non-empty subset $C$ of $X$ and any non-empty subset $V'$ of  $D(\!\supseteq\!C)$, we have $C \subseteq \bigcap_{v\in V'}\overrightarrow v  \subseteq \overrightarrow{V'}=\bigcup_{v\in V'}\overrightarrow v$ as well as
\begin{equation}
\label{basic}
 \overrightarrow{V'}=C 
\, \iff \,V'\subseteq D(\!=\!C).
\end{equation}

Clearly, a subset $C$ of $X$ is a tight cluster if and only if just one subset $V=V_C$ of $D(\!=\!C)$ separates $C$ from $X-C$.

As an example, the non-singleton tight $X$--clusters of the graph $G$ shown in Fig. 2 are $\{a,b\}$ and $X$, as $D(=\{a,b\}) = \{v_1,v_2,v\}$ holds where $v$ is the left-hand parent of $v_1$ and $v_2$, and this set clearly separates $\{a,b\}$ from $\{c,d,e\}$; yet  the subset $\{v_1, v_2\}$ of $D(=\{a,b\})$ also separates $\{a,b\}$ from $\{c,d,e\}$.

Notice that $X$ itself is a tight cluster if and only the strongest connectivity condition (C1) holds. Notice also that
the set of tight $X$---clusters of $G$ is always a subset of the hierarchy $\Hh_\Pc$ defined in (\ref{forfree}) for $\Pc = \{\overrightarrow{v}: v \in V\},$ though, in general,
the latter set can be strictly larger than the set of tight $X$--clusters of $G$.

The concept of a tight cluster is a  relaxation of the notion of `organismic exclusivity' described recently by \cite{bau}, which requires that there is an element in $D(=C)$ that separates
$C$ from $X-C$.

\bigskip

\subsection{An example of a tight cluster in recent evolution}

The conditions for a tight cluster are strong. However, results in population genetics suggest that for diploid
(sexually-reproducing) populations, it may sometimes be reasonable. This is because, under a neutral model of random
diploid mating, \cite{cha} showed that  if we trace back the ancestry of a set of $n$ extant individuals by (at least)
$1.77\log_2(n)$ generations, the population extant at this earlier time is
likely to have the property that each individual in this ancestral population either has no extant descendants,
or has all  $n$ extant individuals as descendants. This sharp $\log_2(n)$ behavior was shown to extend to more realistic models of human mating behaviour, including migration,
at the price of a constant larger than $1.77$ by \cite{roh}.

The significance of this finding can be illustrated by considering, for example, the entire extant human population
$P_{\rm hom}$ as a subset of the set $X$ of all extant organisms on earth today.  The work of \cite{roh}, along with recent evidence that the radiation of
modern humans from Africa occurred within the last 150,000 years \citep{africa}
suggests that -- excluding the existence of a {\em Homo erectus} type Yeti or Bigfoot -- every individual $v$ in the population $V_{\rm hom}$ that was (i) ancestral to $P_{\rm hom}$ and (ii) living (say) 200,000 years ago, satisfies either $\overrightarrow{v} \cap P_{\rm hom}= P_{\rm hom}$, or $\overrightarrow{v}\cap P_{\rm hom}= \emptyset$.  Moreover, we can presumably be confident that no other non-human individual organism alive today is a descendant of any individual in $V_{\rm hom}$ and, so, $V_{\rm hom}$ would satisfy the
conditions for the set $V_C$ mentioned above: it is tight, i.e.
$\overrightarrow{V_{\rm hom}}=P_{\rm hom}$ holds and it separates $P_{\rm hom}$ from all other currently living organisms.

Thus, we may assume that $P_{\rm hom}$ is, formally, a tight cluster in the set $X$ of all extant organisms alive today.

The example also underlines that, because of our specific choice of $X$, side lines with no descendants today (like, presumably, the Neanderthals) are of no direct interest in this context. Indeed, we may probably (that is, unless
the Yeti or Bigfoot exists and belongs to the {\em Homo erectus} group) also take for $V_{\rm hom}$ all individuals that were  ancestral to $P_{\rm hom}$ and lived 2,000,000 years ago, which, however, would not work if we choose $X$ to denote all humans from the last 1,000,000 years that had no children.

In the case of haploid reproduction, coalescence times are much longer, being of order $n$ rather than $\log(n)$ \citep{hei}.  Nevertheless, consider a  current population of $n$  individuals with haploid reproduction.  Suppose the ancestors of this population dating back as far as $N$ generations into the past constituted a homogeneous population was of approximately constant size, and was genetically isolated (i.e. if there were LGT events involving this ancestral population then they were restricted to exchanges between members of that population) and which left no other descendants today. Then, provided $N>>n$, this current population would be a likely candidate for a tight cluster in the set $X$ of all extant organisms.
$\hfill$ $\square$

\subsection{Strict clusters}

We  now describe a second class of clusters; we will see in Theorem~\ref{mainthm} that these include the tight clusters, yet they are still guaranteed to form a hierarchy.

Define  a subset $C$ to be a {\em strict $X$--cluster} (relative to $V$ and $\preceq$) provided that
 \begin{itemize}
 \item
 $v\in V$ and $C\cap \overrightarrow v\neq \emptyset$ implies that either $C\subseteq  \overrightarrow v$ or
$  \overrightarrow v\subseteq  C$ -- or, equivalently, $v\in D(\!\supseteq \!C)$ or $v\in D(\!\subseteq \!C)$ holds, and
\item
the {\em 	cousinship graph}
$$
\Gamma(C):=(C,\big\{\{x,y\}\in \binom{C}{2}: \exists v\in  D(\!\subseteq\! C): x,y\in \overrightarrow v\big\})
$$
of $C$ is connected.
\end{itemize}
As an example, the non-singleton strict $X$--clusters of the graph $G$ shown in Fig. 2 are $\{a,b\}, \{d,e\}$ and $X$.

Notice that $X$ is a strict cluster if and only if the weakest connectivity condition (C3) holds.\\

\subsection{Clusters based on ancestry}

We begin this sub-section with some further definitions.

For any pair of elements $\{a,b\}$ in $V,$ let $${\rm ca}(a,b):= \{v\in V~:~v\preceq a~\text{and}~v\preceq b\}$$ be the set of {\em common ancestors} of $a$ and $b$.  Provided that ${\rm ca}(a,b)$ is non-empty,  let ${\rm mrca}(a,b)$ be the maximal elements in ${\rm ca}(a,b)$; this is often referred to as the set of  the {\em most recent common ancestors} of $a$ and $b$.
For $a,b,c\in X$, let us write $ab||c$ if ${\rm ca}(a,b)$ is non-empty, and for each $v\in {\rm mrca}(a,b)$ there exists $v' \in {\rm mrca}(a,c)$ and $v'' \in {\rm mrca}(b,c)$ such that $v' \prec v$ and $v'' \prec v$ hold.

As an example, for the graph $G$ in Fig. 2, we have $ab||x$ for each $x \in \{c,d,e\}$, and we have $de||y$ precisely when $y \in \{a,b\}$.

We will write $ab|c$ under the strictly weaker condition that  ${\rm ca}(a,b)$ is non-empty, and there exists, for each $v\in {\rm mrca}(a,b)$, some $v' \in {\rm mrca}(a,c) \cup {\rm mrca}(b,c)$ with $v' \prec v$.

A dual notion to the ancestral relation $||$  is the following: For $a,b,c\in X$, let us write $ab \perp c$ if ${\rm ca}(a,c)$ and ${\rm ca}(b,c)$ are both non-empty and there exist, for all $v\in {\rm mrca}(a,c)$ and $v' \in {\rm mrca}(b,c)$, some $u, u' \in {\rm mrca}(a,b)$ (where $u$ need not necessarily be different from $u'$) such that $v \prec u$ and $v' \prec u'$ holds.
Note that $||$ is neither stronger or weaker than $\perp$, that is, there are examples for which $xx' \perp y$ holds but $xx'||y$ fails (Fig. 3(a)) and also for which $xx'||y$ holds but $xx' \perp y$ fails (Fig. 3(b)).

\begin{figure}[ht] \begin{center}
\resizebox{12cm}{!}{
\includegraphics{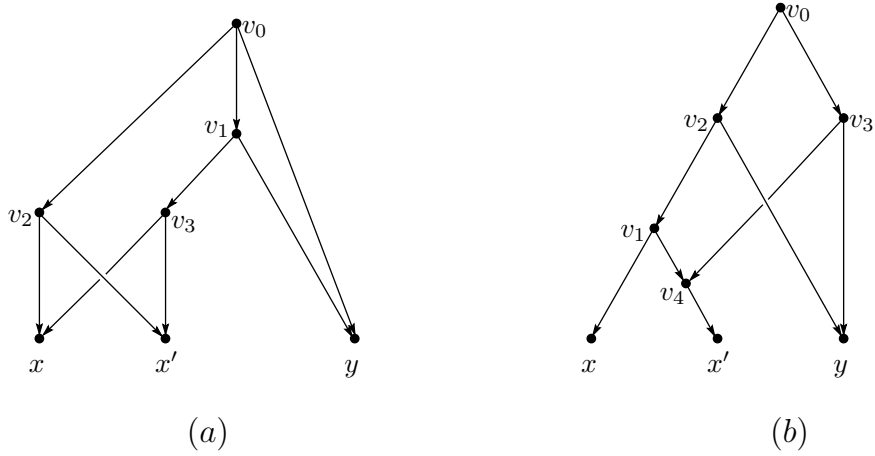}
} \caption{(a) An acyclic digraph $G$ on $X = \{x,x',y\}$ for which $\{x,x'\}$ is a tight cluster and a co-ancestral cluster but is not an  ancestral cluster. (b) An acyclic digraph $G$ on $X=\{x,x',y\}$ for which $\{x,x'\}$ is an ancestral cluster but not a co-ancestral cluster.}
\end{center}
\label{figure2}
\end{figure}

The following result summarizes a basic property of these relations, and will be useful in the next section.
\begin{lemma}
\label{lem:triplet}
Suppose that $G$ is any finite, directed, cycle-free graph, with $X \subseteq V$.  Given three distinct elements $a,b,c\in X$:
\begin{itemize}
\item[\rm (i)] At most one of $ab||c$, $ac||b$ and $bc||a$ holds;
\item[\rm (ii)] At most two of $ab|c, ac|b, bc|a$ holds;
\item[\rm (iii)] At most one of $ab\perp c, ac\perp b, bc \perp a$ holds.
\end{itemize}
\end{lemma}

\noindent
{\em Proof:} For part (i), assume that both $ab||c$ and $ac||b$ hold. Let $v$ be any element in ${\rm mrca}(a,b)$; then there exists $v'\in {\rm mrca}(a,c)$ with $v'\prec v$. On the other hand, there also exists an element $u\in {\rm mrca}(a,b)$ such that $u\prec v'$ in view of $ac||b$. Therefore, we have $u\prec v$ and  $u,v\in {\rm mrca}(a,b)$, a contradiction to the definition of ${\rm mrca}(a,b)$.
The second and third parts follow by a similar proof by contradiction. This completes the proof of the Lemma.
\hfill$\Box$

With these definitions, we say that $C$ is a {\em ancestral $X$--cluster} (respectively  {\em relaxed ancestral $X$--cluster} and {\em  co-ancestral cluster}) if for all $x,x'\in C$ and $y \in X-C$, we have $xx'||y$
(respectively $xx'|y$ and $xx' \perp y$).  Notice that the entire set $X$ is both an ancestral cluster and a co-ancestral cluster under the intermediate connectivity condition (C2).

Note that, even for a digraph $G$ that has a vertex $v_0$ with $\overrightarrow{v_0} = X$, there may exist a tight $X$--cluster that is not an ancestral cluster, as Fig. 3(a) shows for $C = \{x,x'\}$.  In this example,
$D(=C) = \{v_2, v_3\}$, from which it is easily seen that $C$ is a tight cluster. Note that $v_2 \in {\rm mrca}(x,x')$ yet $v_2$ is not a descendant of any vertex in either ${\rm mrca}(x,y)=\{v_1\}$ or ${\rm mrca}(x',y) = \{v_1\}$.

\subsection{Clusters relative to a `time scale'}
In this section, we exploit an additional aspect of evolution -- the fact that the vertices of $G$ have an associated `date' (e.g. time when they were born) and this provides a further avenue to define a system of clusters.

Suppose that, in addition to the digraph $G = (V,A)$, with $X \subseteq V$, we have a map $T:V \rightarrow \Rb$ that strictly preserves the partial order $\preceq$, i.e.
$$u \prec v \Longrightarrow T(u) < T(v).$$ We refer to the pair $(G,T)$ as a {\em valuated digraph on $X$}.
Of course, the condition that such a map $T$ exists is equivalent to the condition that $G$ has no directed cycles \citep{jor}, but we think of $T$ as being a  specific map, where,  in the biological context, $T(v)$ would denote the time when the individual $v$ was born (we may regard the present as time $0$ and so $T$ is a map from $V$ to the non-positive reals).

Following \cite{steel} we  say that $C \subseteq X$ is a {\em Apresjan $X$--cluster relative to $T$} if there exists $t \in \Rb$ such that:
\begin{itemize}
\item[(T1)] For all $x,y \in C$, there exists $v \in V: v \preceq x,y, T(v) \geq t;$ and
\item[(T2)] For all $x \in C$, $y \in X-C$, if $v \in V$ satisfies $v \preceq x,y$ then $T(v) <t$.
\end{itemize}
In words, $C$ is an Apresjan $X$--cluster relative to $T$ if every two individuals in $C$ have at least one common ancestor after time $t$,
but each individual in $C$ and each individual in $X-C$
have all their common ancestors earlier than $t$.

We say that $C \subseteq X$ is a {\em strong Apresjan $X$--cluster relative to $T$} if
(T1) is strengthened to:
\begin{itemize}
\item[(T1$'$)] For all $x,y \in C$, and {\em every} $v \in {\rm mrca}(x,y)$, $T(v) \geq t$.
\end{itemize}
Thus, $C$ is a strong Apresjan $X$--cluster relative to $T$ if every two individuals in $C$ have {\em all} their
most recent common ancestors after time $t$, but any individual in $C$ and individual in $X-C$ have all their
common ancestors earlier than $t$.

\section{{\sc Main result}}
We have described a variety of ways to construct a set of $X$--clusters from $G$. We now show that they all lead to hierarchies (in one case a weak hierarchy),
and describe some relationships between them, in the following main result of this paper, the proof of which is given in the Appendix.

\begin{theorem}
\label{mainthm}
Suppose that $G$ is any finite, directed, cycle-free graph, with $X \subseteq V$.
\begin{enumerate}
\item
The following sets form a hierarchy:
\begin{enumerate}
\item The set of tight $X$--clusters of $G$;
\item The set of strict $X$--clusters of $G$;
\item The set of ancestral $X$--clusters of $G$;
\item The set of  co-ancestral $X$--clusters of $G$.
\end{enumerate}
\item
The set of relaxed ancestral $X$--clusters of $G$ forms a weak hierarchy.
\item
Suppose that $(G,T)$ is a valuated digraph on $X$. Then the set of Apresjan $X$--clusters relative to $T$ forms a hierarchy (as does the the subset of strong Apresjan $X$--clusters relative to $T$).
\item
Every tight $X$--cluster $C$ of $G$ is also a strict $X$--cluster and, under connectivity condition {\rm (C2)}, a co-ancestral cluster.  If $G$ has a valuation map $T$, $C$ is also an  Apresjan $X$--cluster relative to $T$.
\end{enumerate}
\end{theorem}

\section{{\sc Discussion}}

Our paper is motivated partly as a response to a currently promoted  viewpoint that processes of reticulate evolution, such as extensive LGT implies that no sensible or well-defined `tree of life' can be constructed  \citep{doo, koo, law}.
However, this statement depends on  how  one views such a tree, and where the transfer events occurred in it.  For example, even if each gene has been transferred once during its history \citep{dag}, provided that these transfer events all occurred before the separation of certain populations then we may still expect to find Apresjan or stronger (e.g. tight) clusters, which will therefore form a tree. Consider,  for example, the collection $C$ of all extant mammals. The most recent common ancestors of mammals most likely occurred within the last 120 million years \citep{elz}. Thus if those genes that are found in mammals and which underwent a gene transfer event some time in their past did so at a much earlier stage of evolution (i.e. well before 120 million years ago) then the concept of a `mammal tree' composed of clusters of a type described above seems reasonable.

Neither are recent LGT events necessarily problematic. In particular, such events  will not destroy even a tight cluster $C$ provided they occur amongst those ancestors of $C$ that are descendants of $D(=C)$.

For prokaryotes, where a tree structure is most vigorously called into question, the concept of a tree is still well defined, but it may indeed be poorly resolved (depending on the type of cluster considered, and the extent to which a LGT event from individual $x$ to $y$ might be counted as an arc in $G$ from $x$ to $y$ -- for example, one could indicate all such instances or just those for which the gene transfers survives to a present copy).  In cases where  LGT (and other types of reticulate evolution) are extensive and on-going, then set systems such as weak hierarchies may give a more informative picture of evolution than a tree. We have described one way to generate such a hierarchy above, but it may be useful to explore other approaches.

\begin{figure}[ht]
\begin{center}
\resizebox{6cm}{!}{
\includegraphics{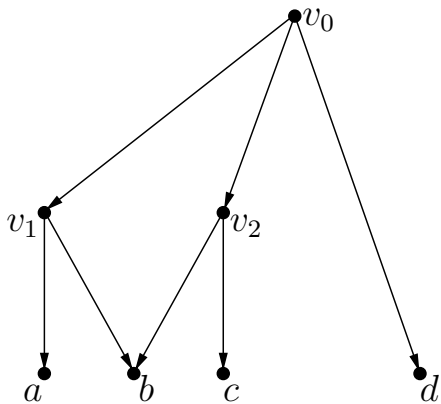}}
\caption{An example to illustrate a violation of sampling consistency for strict clusters.}
\end{center}
\label{figure1}
\end{figure}

In this paper, we have concentrated instead on ways by which a hierarchy on $X$ can be constructed from $G$ based on concepts of ancestry and separation.  Of course, the possibilities we have outlined are by no means exhaustive, as there will surely be other combinations of conditions that will allow for a hierarchy or related set system.  However, we would like any procedure for constructing a hierarchy to have some reasonable biological motivation
and also, if possible, to satisfy some desirable properties.
One such desirable property is that the procedure be `robust' with respect to the possibility that we have not sampled or
observed all individuals in $X$.  We can make this precise as follows.

Suppose that $G$ is any finite, directed, cycle-free graph with $X \subseteq V$, and  that $Y$ is a subset of $X$.  Let $G|Y$ be the directed graph obtained from $G$ by
regarding the vertices in $X-Y$ as unlabeled vertices.
Now suppose we have a function $\phi$ that associates to each such pair $(X,G)$ a collection of subsets of $X$.
We say that $\phi$ satisfies {\em sampling consistency} if it satisfies the condition:
$$C \in \phi(X,G) \Rightarrow C \cap Y \in \phi(Y, G|Y).$$  We can extend this concept to valuated digraphs in the obvious way (namely,  $C \in \phi(X,G,T) \Rightarrow C \cap Y \in \phi(Y,G|Y,T)$).

It can be checked that the following constructions satisfy sampling consistency: tight clusters,
ancestral clusters, and Apresjan clusters (with respect to a time scale). However, the strict cluster construction
can violate this condition -- for example, consider the graph $G$ in Fig. 4. Then $C=\{a,b,c\}$ is a strict
$X$--cluster where $X= \{a,b,c,d\}$. But if we select $Y= \{a,c,d\}$ then $C \cap Y = \{a,c\}$ is not a strict $Y$
cluster in the graph $G|Y$, since the cousinship graph $\Gamma(C \cap Y)$ is not connected.

\section*{{\sc Acknowledgements}}
{\em V.M. and T.W. thank the Engineering and Physical Sciences Research Council (EPSRC) for its support [Grant EP/D068800/1].}

\bibliographystyle{sysbio}
\bibliography{species}

\section{{\sc Appendix: Proof of Theorem 1.}}

\noindent {\em Proof of Part 1(a):} Suppose that for two tight clusters $C_1$ and $C_2$ we have $C_1 \cap C_2 \neq \emptyset$ and that $C_2$ is not a subset of $C_1$. We will show that $C_1 \subseteq C_2$.
Let $V_1 = D(=C_1)$ and $V_2=D(=C_2)$.
By assumption, there exists $x \in C_1 \cap C_2$, $y \in C_2-C_1$.  First observe that if $V_2 \subseteq V_1$ then
$\overrightarrow{V_2} \subseteq \overrightarrow{V_1}$, which implies that $C_2 \subseteq C_1$ in violation of our assumption.
Thus, there exists $v \in V_2- V_1$.  Now, since $x,y \in C_2$, and  $v \in V_2$ there exists a directed path from $v$ to $x$ and a directed path from $v$ to $y$. In particular these provide an undirected path $P$ in $G$ connecting $x$ and $y$. But now, since $x \in C_1$ while $y \in X-C_1$, and since $C_1$ is tight (so $V_1$ separates $C_1$ from $X-C_1$)  at least one vertex, say $w$, in $P$
must lie in $V_1$.  Regardless of where $w$ lies on $P$ we have $v \preceq w$ (since every vertex $v'$ on $P$ satisfies $v \preceq v'$) and so $\overrightarrow  w \subseteq \overrightarrow  v$. Therefore, since
	$\overrightarrow w = C_1$ and $\overrightarrow  v = C_2$, we have $C_1 \subseteq C_2$, as required. This completes the proof of   {\em Part 1(a)}.

To establish  {\em Part 1(b)}, suppose that $C,C'$ are strict $X$--clusters, and that $C \cap C'$ and $C-C'$ are both non-empty. We will show $C' \subseteq C$.  Take $x \in C \cap C', y \in C-C'$.  By the
connectivity of the cousinship graph $\Gamma(C)$ there is a path in this graph from $x$ to $y$, say $x = x_1, x_2, \ldots, x_k=y$.  Let $x_i, x_{i+1}$ be the first pair of adjacent vertices in this path for which $x_i \in C \cap C'$ and $x_{i+1} \in C-C'$.
Since $x_i$ and $x_{i+1}$ are adjacent there is a vertex $v \in V$ for which $x_i, x_{i+1} \in \overrightarrow v \subseteq C$.  Moreover, we have $\overrightarrow v \cap C' \neq \emptyset$ (since
$x_{i}  \in C' \cap \overrightarrow v$) and so  the first condition in the definition of a strict cluster implies that either $C' \subseteq \overrightarrow v$ or $\overrightarrow v \subseteq C'$.  But
the second of these two inclusions is impossible, since $x_{i+1}  \in \overrightarrow v-C'$.  Thus $C' \subseteq \overrightarrow v$ and since $\overrightarrow v \subseteq C$, this implies
that $C' \subseteq C$, as required to establish  {\em Part 1(b)}.

For {\em Part 1(c)}, assume, for the sake of contradiction, that $C,C'$ are ancestral clusters, and there exist three elements
$a,b,c$ with $a\in C-C'$, $b\in C'-C$ and $c\in C\cap C'$.
Then, by definition, we have $ac||b$ and $bc||a$, a contradiction to Lemma~\ref{lem:triplet}(i).
	A similar argument applies for  {\em Part 1(d)}. This completes the proof of  {\em Part 1}.

\noindent {\em Proof of Part 2:}
Suppose that $A,B,C$ are three relaxed ancestral clusters which violate the condition $A \cap B \cap C \not\in \{A \cap B, A \cap C, B \cap C\}$.  Then we can select $x \in A\cap B -  C, y \in A \cap C -  B, z \in B\cap C - A.$
We have $xy|z$ (since $x$ and $y$ but not $z$ are in $A$), and $xz|y$ (since $x$ and $z$ but not $y$ are in $B$), and $yz|x$
(since $y$ and $z$ but not $x$ are in $C$), in violation of  Lemma~\ref {lem:triplet}(ii).

\noindent {\em Proof of Part 3:}
This result is from \cite{steel}, based on  earlier related results from \citep{apr, bry, dev}. Since the proof is short, we provide it here for completeness.
Suppose $C_1, C_2$ are Apresjan $X$--clusters relative to $T$ and there exists $x \in C_1 \cap C_2, y \in C_1-C_2, z \in C_2-C_1$;  we will show that this leads to a contradiction.  For $i \in \{1,2\}$, let  $t_i$ be a value of $t$ for which (T1), (T2) applies for $C=C_i$.   If $t_1 \geq t_2$ then, by condition (T1) on $C_1$, there exists $v$ with $v \preceq x, y$ with $T(v)  \geq t_1 \geq t_2$. But  applying (T2) to $C_2$  gives  $T(v) < t_2$ (since $y \in X-C_2$), a contradiction.
A similar argument applies if $t_1 \leq t_2$.

\noindent  {\em Proof of Part 4:}
Suppose that $C$ is a tight $X$--cluster. We first show that $C$ is a strict $X$--cluster.  Select any $w \in D(=C)$.  Then $\overrightarrow{w} =C$, and
so the cousinship graph $\Gamma(C)$ is a clique (and hence a connected graph).   Now, suppose that $C \cap \overrightarrow{v} \neq \emptyset$, and that $\overrightarrow{v}$ is not a subset of $C$.
We will show that $C \subseteq \overrightarrow{v}$.  Select $x \in C \cap \overrightarrow{v}, y \in \overrightarrow{v} -C$.  There exists a directed path in $G$ from $v$ to $x$ and a directed path from $v$ to $y$. In particular, these provide an undirected path $P$ in $G$ connecting $x$ and $y$.
Since $x \in C$ but $y$ lies outside of $C$, path $P$ must contain at least one vertex $v' \in D(=C)$ (since $D(=C)$ separates $C$ from $X-C$).  Then $v \preceq v'$ and so $\overrightarrow{v'} \subseteq \overrightarrow{v}$. But
 $\overrightarrow{v'} = C$ (since $v' \in V_C$) so that $C \subseteq \overrightarrow{v}$, as required to establish that $C$ is strict $X$--cluster.

 Next we show that $C$ is a co-ancestral cluster, i.e.  for any $x,x' \in C, y \in X-C$ we have $xx'\perp y$.  Let $v$ be a vertex in
 ${\rm mrca}(x,y)$ (such a vertex exists by (C2)) and consider the (undirected) path $P$ from $x$ to $v$ to $y$.  Since $x \in C$ to $y \in X-C$, the fact that $D(=C)$ separates $C$ from $X-C$ (because $C$ is a tight cluster) implies that one vertex, say $w$, in $P$  must lie in $D(=C)$. The vertex $w$ does not lie on the path from $v$ to $y$, otherwise we have $y \in \overrightarrow{v} = C$, so $w$ is in the path from  $v$ to $x$. Since
 $x' \in \overrightarrow{w}$, it follows that $w$ is, or has as a descendant, a vertex in ${\rm mrca}(x,x')$.
 A similar argument applies to any vertex in ${\rm mrca}(x', y)$ and so $xx'\perp y$. Since this holds for all $x,x' \in C$ and $y \in X-C$, $C$ is a co-ancestral cluster of $X$.

For the final claim in {\em Part 4}, suppose that $C$ is a tight $X$--cluster of $G$.
We will show that (T1) and (T2) hold for $t=t_C$ where:
$t_C:= \max\{t(v): v \in D(=C)\}.$
First select $v_0 \in V_C$ with $T(v_0) = t_C$.
Observe that for all $x,x' \in C$, we have $v_0 \preceq x,x'$ and since $T(v_0) \geq t_C$ we see that condition (T1) is satisfied for $t= t_C$, and $v = v_0$.  To verify condition (T2), suppose that $x\in C, y \in X-C$ and there exists $v \preceq x,y$ with $T(v) \geq t_C$.  Consider the (undirected) path $P$ in $G$ from $x$ to $v$ and then to $y$. If $v \in V_C$ then $\overrightarrow{v} = C$ which is impossible since
$v \preceq y \Rightarrow y \in \overrightarrow{v}$ yet $y$ is not an element of $C$. Thus $v$ is not an element of $V_C$.  Moreover, for any vertex $w$ in $P$ that is different from $v$, we have $T(w) > t_C$ (since $v \prec w$ and $T$ is strictly monotone) and so $w$ is also not an element of $V_C$
(since all the vertices $w'$ in $V_C$ satisfy
$T(w') \leq t_C$). In summary, none of the vertices in $P$ belongs to $V_C$, thus deleting $V_C$ fails to disconnect $x$ from $y$, violating the assumption that $D(=C)$ separates $C$ from $X-C$.  This establishes property (T2), as required, and thereby completes the proof.
$\hfill$ $\square$

\end{document}